# Analytical phase optical transfer function for Gaussian illumination and the optimized profiles

**JIANHUI HUANG,[1,2] YIJUN BAO,[3] AND THOMAS K. GAYLORD[2]**

[1]*School of Mechanical Engineering, Shanghai Jiao Tong University, Shanghai, 200240, China.*
[2]*Optics Laboratory, School of Electrical and Computer Engineering, Georgia Institute of Technology, Atlanta, Georgia 30332-0250, USA*
[3]*Department of Biomedical Engineering, Duke University, Durham, NC 27708, USA*

*\*tgaylord@ece.gatech.edu*

**Abstract:** The imaging performance of tomographic deconvolution phase microscopy can be described in terms of the phase optical transfer function (POTF) which, in turn, depends on the illumination profile. To facilitate the optimization of the illumination profile, an analytical calculation method based on polynomial fitting is developed to describe the POTF for general non-uniform axially-symmetric illumination. This is then applied to Gaussian and related profiles. Compared to numerical integration methods that integrate over a series of annuli, the present analytical method is much faster and is equally accurate. Further, a "balanced distribution" criterion for the POTF and a least-squares minimization are presented to optimize the uniformity of the POTF. An optimum general profile is found analytically by "relaxed optimal search" and an optimum Gaussian profile is found through a tree search. Numerical simulations confirm the performance of these optimum profiles and support the "balanced distribution" criterion introduced.

## 1. INTRODUCTION

Three-dimensional quantitative phase imaging (3D QPI) techniques using partially coherent (PC) light sources such as halogen lamps, have been implemented on conventional bright-field microscope platforms. Described as partially coherent optical diffraction tomography (PC-ODT), this technology has revealed the internal microstructure of transparent objects such as cells and optical fibers [1-6]. The object's structure is reconstructed from stacks of 2D defocused images taken at multiple illumination angles. Unlike holographic tomographic 3D QPI that applies intermediate 2D phase retrieval to reconstruct the 3D volume refractive index (RI) distribution, PC-ODT applies volume diffraction theory to recover directly the structure of the phase object. Rather than using 2D phase retrieval, the first-order Born approximation is assumed to allow linear reconstruction of the total object [7-11]. Recently, a multi-layer multiple-scattering light propagation model has been proposed to improve the accuracy of PC-ODT for strongly scattering objects [12, 13]. Due to the use of partially coherent illumination, PC-ODT can achieve, at most, twice the transverse resolution of coherent ODT. To achieve this limiting resolution, researchers need to use the maximum NA, but a simple disk illumination with maximum NA cannot produce any contrast, so a better illumination profile is needed to optimize the performance. This open question is addressed and answered in the present work. Generally, the imaging performance of PC-ODT can be well predicted from the phase optical transfer function (POTF) of the optical system. The POTF is a 3D function in spatial frequency coordinates that gives the weighting of the object's spatial frequency components in producing the diffracted intensities. Thus, in the present work, the imaging performance of PC-ODT is enhanced by introducing a "balanced distribution" criterion for the POTF distribution. This, in turn, is achieved by finding the most suitable illumination profile.

Starting from the Zernike-phase-contrast microscopy that applied thin-annular illumination rather than disk illumination [14], the selection and optimization of the illumination format has been the subject of continuous interest. A typical example of this is the 2D QPI based on the transport-of-intensity equation (TIE). Initially, the TIE method was only applicable for coherent phase imaging [15, 16]. After several decades, the meaning of the "phase" has been extended to include partially coherent light and PC-TIE methods were developed [17-20]. In general, increasing the numerical aperture of the illumination ($NA_{illu}$) is expected to increase the resolution limit. However, simultaneously, the phase contrast effect is degraded when $NA_{illu}$ increases, and this reduces the resolution [21]. By replacing disk illumination with annular illumination, Zuo has achieved a resolution near the incoherent diffraction limit of 208nm and an effective $NA$ of 2.66 [22]. Others have also reported that annular illumination can be more advantageous than disk illumination in enhancing resolution [23-29]. These results represent an important beginning in the pursuit of illumination optimization for 3D PC-ODT. Due to the complex relationship between the illumination profile and the resulting POTF, current investigations have basically been empirical, rather than based on a theoretical foundation. For example, from experimental results, Jenkins observed that Gaussian illumination produces resolution-enhanced RI recovery with fewer artifacts compared with uniform illumination [9]. Also, by comparing the distributions of the POTF values, Soto showed that Gaussian illumination was superior to disk illumination in terms of resolution and accuracy [30]. Related approaches that synthesize improved POTFs to achieve high-resolution imaging are the application of complementary illuminations and the use of multiple illumination profiles in sequence [10, 31]. However, the time-consuming sequence of intensity measurements makes that approach unusable for real-time 3D QPI. When imaging frame rate is a consideration, optimizing the illumination becomes a logical approach for enhancing the resolution of existing PC-ODT systems. Up to the present time, there has been a lack of systematic research to provide a theoretical foundation for evaluating the effects of the illumination profile. A general tool is needed to enable determining optimal illumination profiles based on tailoring the POTF distribution.

In this paper, a "balanced distribution" criterion is proposed for determining the optimal POTF. Using this criterion as the optimization target, the magnitude of the resulting POTF is more uniformly distributed among high and low spatial frequencies. A least square method is used to evaluate quantitatively the POTF. By relaxing the criterion, an optimal constraint-free illumination profile named "relaxed optimal illumination" is calculated analytically. In addition, to determine the optimum Gaussian profile, this paper develops an analytical calculation method for the Gaussian POTF based on polynomial fitting, which enables the non-paraxial POTF to be determined for general non-uniform axially-symmetric illuminations. Previously, an analytical POTF expression was available only for uniform illumination [8]. Now, with the general formulation available, an optimum Gaussian profile is found quickly through a tree search. Numerical simulations confirm the performance of these optimum profiles and support the "balanced distribution" criterion introduced in this work.

## 2. CALCULATION OF POTF FOR GAUSSIAN ILLUMINATION

Gaussian illumination is widely used in 3D QPI. It produces a generally favourable distribution within the POTF. High-resolution PC-ODT often uses equal numerical apertures of the condenser and objective lenses, but in this case, the uniform illumination yields an all-zero POTF and is therefore completely unable to image the phase. By comparison, general Gaussian illumination gives a non-zero POTF and produces useful phase imaging. The details of the imaging performance are highly correlated with the POTF distribution [30]. To be able to predict imaging performance and to pursue illumination optimization, it is important to develop a general non-paraxial analytic representation for Gaussian and other nonuniform illuminations. Bao presented the concept of using the analytical POTFs of thin-annular rings of illumination to calculate the POTF of non-uniform illuminations including the case of Gaussian illumination [8]. In this section, two approaches to calculate the Gaussian POTF are presented. The first is similar to Bao's method but divides the Gaussian profile into uniform disks instead rather than annuli. The second is based on fitting the Gaussian profile by polynomials, from which an analytical POTF formula is then derived.

A. Gaussian POTF by Segment Division

From pioneering work of Streibl [32], the phase optical transfer function (POTF) is a function defined on the spatial frequency domain ($\eta$, $\mu$), where $\eta$ indicates the object's axial spatial frequency and $\mu$ the lateral spatial frequency. Generally, the POTF is determined by the objective function $O$ and the specific illumination function $S$. The integration formula for the non-paraxial POTF is given below.

$$T(\eta,\mu) = \int O(\mathbf{k}'+\mathbf{\mu}/2) O(\mathbf{k}'-\mathbf{\mu}/2) \left[ S(\mathbf{k}'+\mathbf{\mu}/2) - S(\mathbf{k}'-\mathbf{\mu}/2) \right] \\ \cdot \delta\left( \sqrt{1-(\mathbf{k}'+\mathbf{\mu}/2)^2} - \sqrt{1-(\mathbf{k}'-\mathbf{\mu}/2)^2} - \eta \right) d\mathbf{k}'. \tag{1}$$

In previous research, the general POTF formula for uniform illumination was given by Eq. (8) in [8] and Eq. (19) in [33]. A simplified analytical version of these is

$$T(\eta,\mu) = \frac{2(1-a)b + \eta^2 a}{2\eta a \sqrt{a}} (\theta_D - \theta_S) + \frac{(1-a)b}{2\eta a \sqrt{a}} (\sin 2\theta_D - \sin 2\theta_S), \tag{2}$$

where

$$\begin{cases} a = 1 + \dfrac{\mu^2}{\eta^2}, \quad b = 1 - \dfrac{\mu^2 + \eta^2}{4} \\ \theta_D = \arcsin \sqrt{\dfrac{b - a x_D^2}{b}}, \quad \theta_S = \arcsin \sqrt{\dfrac{b - a x_S^2}{b}} \\ x_D = \dfrac{\eta}{2\mu} \cdot \left\{ -\eta - 2\sqrt{1-\rho_O^2}, \ \eta - 2\sqrt{1-\rho_C^2} \right\}_{\min} \\ x_S = \dfrac{\eta}{2\mu} \cdot \left( -\eta - 2\sqrt{1-\rho_C^2} \right) \end{cases} \tag{3}$$

The above Eqs. (2) and (3) establish the analytical POTF solution for uniform disk illumination that are used in the presented work. This POTF will be represented by $T_U \langle \rho_O, \rho_C \rangle$, where $\rho_O$ and $\rho_C$ are the normalized numerical apertures (NAs) of the objective and the condenser, respectively [33]. Any non-uniform axially symmetric illumination including Gaussian, can be accurately approximated by a series of thin annular rings (or thin discs) of varying intensities. Since the POTF integration of Eq. (1) is linear with respect to illumination intensity, the POTF for a given axially symmetric illumination $S(k)$ is also a linear combination of uniform disk illumination POTFs.

$$T = \sum_{n=1}^{N} \left[ S\left(\frac{n}{N}\rho_C\right) - S\left(\frac{n+1}{N}\rho_C\right) \right] \cdot T_U \left\langle \rho_O, \frac{n}{N}\rho_C \right\rangle. \tag{4}$$

In Eq. (4), the step size is $\rho_C/N$, the interval for illumination is [0, $\rho_C$], and $S(k)$ is 0 when $k > \rho_C$. The accuracy of Eq. (4) is affected by the step size. To obtain an accurate POTF for Gaussian illumination, $N$ needs to be sufficiently large. In the present research, the calculation of the Gaussian POTF is converted to the calculation of $N$ element POTFs. Therefore, there will be a trade-off between the accuracy and the speed of the calculation.

B. Gaussian POTF by Polynomial Fitting

For the common case of the optical configuration and the illumination both being axially symmetric, the POTF integration expression Eq. (1) can be simplified into a 1D integration expression as shown below. This 1D POTF integral has been established in our previous research as Eq. (17) in [33].

$$T(\eta,\mu) = \int_0^{l_D} \left( \frac{\eta}{2a}\sqrt{\frac{a}{b-k_z^2}} - \frac{2\mu^2}{\eta^3 a}\sqrt{\frac{b-k_z^2}{a}} \right) \cdot S\left( \sqrt{\left(\sqrt{\frac{b-k_z^2}{a}} - \frac{\mu}{2}\right)^2 + k_z^2} \right) dk_z$$
$$- \int_0^{l_S} \left( \frac{\eta}{2a}\sqrt{\frac{a}{b-k_z^2}} - \frac{2\mu^2}{\eta^3 a}\sqrt{\frac{b-k_z^2}{a}} \right) \cdot S\left( \sqrt{\left(\sqrt{\frac{b-k_z^2}{a}} + \frac{\mu}{2}\right)^2 + k_z^2} \right) dk_z. \quad (5)$$

where for Gaussian illumination

$$S(k) = \begin{cases} e^{\sigma k^2}, & k \leq \rho_C \\ 0, & k > \rho_C \end{cases}, \quad (6)$$

where $\sigma$ is the Gaussian width and $k$ is the wavevector magnitude of the incident lightwave ($k_x$, $k_z$), $k = (k_x^2 + k_z^2)^{1/2}$. Due to the Gaussian function $\exp(\sigma k^2)$ in the illumination $S(k)$, it is not possible to achieve a direct analytical solution for the POTF integration when Eq. (6) is substituted into Eq. (5). After a thorough investigation of the integration, it is determined that the analytical solution of POTF for nonuniform illumination exists only when the illumination function $S$ is of a polynomial form. Using this, the Gaussian illumination profile can be approximated by a series of polynomial functions. Therefore, the Gaussian POTF is calculated as a summation of these analytical polynomial POTFs. The procedure is similar to a Taylor expansion and the accuracy depends on the polynomial order. This study applies an 8$^{th}$-order polynomial to fitting the Gaussian illumination profile, i.e.,

$$S(k) = e^{\sigma k^2} \approx p_0 + p_1 k^2 + p_2 k^4 + p_3 k^6 + p_4 k^8 = \sum_{j=0}^{4} p_j (k^2)^j \equiv S^*(k), \quad (7)$$

Where the coefficients $\{p_0, p_1, p_2, p_3, p_4\}$ are determined via the least square method. Then substituting the polynomial-fit Gaussian illumination function $S^*(k)$ of Eq. (7) into the POTF expression Eq. (5) gives

$$T(\eta,\mu) = \sum_{j=0}^{4} p_j \int_0^{l_D} \left( \frac{\eta}{2a}\sqrt{\frac{a}{b-k_z^2}} - \frac{2\mu^2}{\eta^3 a}\sqrt{\frac{b-k_z^2}{a}} \right) \left[ \left( \sqrt{\frac{b-k_z^2}{a}} - \frac{\mu}{2} \right)^2 + k_z^2 \right]^j dk_z$$
$$- \sum_{j=0}^{4} p_j \int_0^{l_S} \left( \frac{\eta}{2a}\sqrt{\frac{a}{b-k_z^2}} - \frac{2\mu^2}{\eta^3 a}\sqrt{\frac{b-k_z^2}{a}} \right) \left[ \left( \sqrt{\frac{b-k_z^2}{a}} + \frac{\mu}{2} \right)^2 + k_z^2 \right]^j dk_z. \quad (8)$$

This equation can be solved in a series of analytical forms. After rather extensive tedious calculations, the final analytical solution for Eq. (8) is given in Table 1. The algorithm is also applicable to lower-order polynomial-fittings. To summarize, the calculation procedure consists of five steps:

(1) Calculate the constant coefficient matrix $h_{p,j}$ from the given coordinate system of $\mu$ and $\eta$.
(2) Approximate the given illumination function $S(k)$ with 8$^{th}$-order polynomial using the least square method and obtain the polynomial coefficients $\{p_0, p_1, p_2, p_3, p_4\}$.

(3) Solve for the integration limits $\{l_D, l_S\}$ through the equations given in the upper right of Table 1.
(4) Substitute $\{l_D, l_S\}$ into the equations given in the middle of Table 1 to obtain $\{U, V\}$.
(5) Put $p$, $U$, $V$ and $h$ into the formula of $G$ and then obtain the analytical POTF solution.

**Table 1.** The analytical POTF formula for rotationally-symmetric illumination

| Analytic POTF Formula for Axially Symmetric Illumination | Solution of ($l_D$ & $l_S$) |
|---|---|
| **Polynomial Fitting** $S(k) \approx p_0 + p_1 k^2 + p_2 k^4 + p_3 k^6 + p_4 k^8$ **POTF Formula** $T(\eta,\mu) = \dfrac{i}{4\pi}\sum_{j=0}^{8}\left[(-1)^j G_j(l_D) - G_j(l_S)\right]$ | $G_j(y) = H_j\left[\dfrac{\eta}{2\sqrt{a}}V_j(y) - \dfrac{2\mu^2}{\eta^3 a\sqrt{a}}U_j(y)\right]$ $H_j = a^{-\tfrac{j}{2}}\sum_{m=0}^{4} p_m h_{p_m,j}$ $a = 1+\dfrac{\mu^2}{\eta^2}\quad b = 1-\dfrac{\mu^2+\eta^2}{4}\quad c = \dfrac{b}{a}$ | $\begin{pmatrix}l_D\\l_S\end{pmatrix} = \sqrt{b - a\begin{pmatrix}x_D\\x_S\end{pmatrix}^2}$ $\begin{pmatrix}x_D\\x_S\end{pmatrix} = -\dfrac{\eta}{\mu}\left(\left\{\dfrac{\eta}{2}+\sqrt{1-\rho_O^2},\ \sqrt{1-\rho_C^2}-\dfrac{\eta}{2}\right\}_{min}\ \dfrac{\eta}{2}+\sqrt{1-\rho_C^2}\right)$ |

| Solution of ($U$ & $V$) $\quad \theta = \arctan(y/\sqrt{b})$ ||
| $j$ | $U_j(y)$ | $V_j(y)$ |
|---|---|---|
| 0 | $b(2\theta + \sin 2\theta)/4$ | $\theta$ |
| 1 | $by - y^3/3$ | $y$ |
| 2 | $b^2(4\theta - \sin 4\theta)/32$ | $b(2\theta + \sin 2\theta)/4$ |
| 3 | $by^3/3 - y^5/5$ | $y^3/3$ |
| 4 | $b^3(12\theta - 3\sin 2\theta - 3\sin 4\theta + \sin 6\theta)/192$ | $b^2(4\theta - \sin 4\theta)/32$ |
| 5 | $by^5/5 - y^7/7$ | $y^5/5$ |
| 6 | $b^3(120\theta - 48\sin 2\theta - 24\sin 4\theta + 16\sin 6\theta - 3\sin 8\theta)/3072$ | $b^3(12\theta - 3\sin 2\theta - 3\sin 4\theta + \sin 6\theta)/192$ |
| 7 | $by^7/7 - y^9/9$ | $y^7/7$ |
| 8 | $b^5(840\theta - 420\sin 2\theta - 120\sin 4\theta + 130\sin 6\theta - 45\sin 8\theta + 6\sin 10\theta)/30720$ | $b^4(840\theta - 672\sin 2\theta + 168\sin 4\theta - 32\sin 6\theta + 3\sin 8\theta)/3072$ |

| Solution of ($h$) |||||
| $j$ | $h_{p_0,j}$ | $h_{p_1,j}$ | $h_{p_2,j}$ | $h_{p_3,j}$ | $h_{p_4,j}$ |
|---|---|---|---|---|---|
| 0 | 1 | $c + \tfrac{1}{4}\mu^2$ | $c^2 + \tfrac{3}{2}c\mu^2 + \tfrac{1}{16}\mu^4$ | $c^3 + \tfrac{15}{4}c^2\mu^2 + \tfrac{15}{16}c\mu^4 + \tfrac{1}{64}\mu^6$ | $c^4 + 7c^3\mu^2 + \tfrac{35}{8}c^2\mu^4 + \tfrac{7}{16}c\mu^6 + \tfrac{1}{256}\mu^8$ |
| 1 | | $\mu$ | $2c\mu + \tfrac{1}{2}\mu^3$ | $3c^2\mu + \tfrac{5}{2}c\mu^3 + \tfrac{3}{16}\mu^5$ | $4c^3\mu + 7c^2\mu^3 + \tfrac{7}{4}c\mu^5 + \tfrac{1}{16}\mu^7$ |
| 2 | | $(a-1)$ | $2(a-1)c + \tfrac{1}{2}(a-3)\mu^2$ | $3(a-1)c^2 + \tfrac{3}{2}(3a-5)c\mu^2 + \tfrac{3}{16}(a-5)\mu^4$ | $4(a-1)c^3 + 3(5a-7)c^2\mu^2 + \tfrac{5}{4}(3a-7)c\mu^4 + \tfrac{1}{16}(a-7)\mu^6$ |
| 3 | | | $2(a-1)\mu$ | $6(a-1)c\mu + \tfrac{1}{2}(3a-5)\mu^3$ | $12(a-1)c^2\mu + 2(5a-7)c\mu^3 + \tfrac{1}{4}(3a-7)\mu^5$ |
| 4 | | | $(a-1)^2$ | $3(a-1)^2 c + \tfrac{3}{4}(a-1)(a-5)\mu^2$ | $6(a-1)^2 c^2 + 3(a-1)(3a-7)c\mu^2 + \tfrac{1}{8}(3a^2-30a+35)\mu^4$ |
| 5 | | | | $3(a-1)^2\mu$ | $12(a-1)^2 c\mu + \tfrac{1}{8}(3a^2-10a+7)\mu^3$ |
| 6 | | | | $(a-1)^3$ | $4(a-1)^3 c + \tfrac{1}{8}(a^3-9a^2+15a-7)\mu^2$ |
| 7 | | | | | $4(a-1)^3\mu$ |
| 8 | | | | | $(a-1)^4$ |

## C. Accuracy and Speed

Given the two calculation methods for the Gaussian POTF, a numerical comparison can be performed to investigate the accuracy and computational speed. As an example, the optical configuration and the Gaussian illumination are set as $\{\rho_O, \rho_C, \sigma\} = \{0.8, 0.7, -2\}$. According to the geometrical symmetry of the POTF, only a quarter of the coordinate system needs to be evaluated, i.e., $\eta > 0$, $\mu > 0$. It is seen that the allowed maximum lateral and axial spatial frequencies of the object are $\mu_{max} = \rho_O + \rho_C = 1.5$, $\eta_{max} = 1- (1-\rho_O^2)^{1/2} = 0.4$, respectively [8]. Therefore, the calculational range of the coordinate system $(\eta, \mu)$ can be set to $\eta < 0.4$, $\mu < 1.5$ to reduce the computing time. To ensure the accuracy of the evaluation, the POTF is calculated on a dense $500 \times 500$ $(\eta, \mu)$ grid. The normalized mean square error (NMSE) may be used to quantify the calculation accuracy. It is defined as

$$NMSE = \frac{\sum (T - T_{ideal})^2}{\sum T_{ideal}^2}. \qquad (9)$$

where $T_{ideal}$ is the ideal POTF. It is unknown but can be well approximated by using the numerical integration method with an extremely large number of segments $N = 10^7$ in Eq. (4). $T$ is the POTF calculated by either the analytical method or the numerical integration method. The simulation is executed in MATLAB 2018b on a laptop with 1.6 GHz 4-core CPU. The results are listed in Table 2.

**Table 2.** Comparison of the POTF's calculation methods ($\{\rho_O, \rho_C, \sigma\}=\{0.8, 0.7, -2\}$)

| Settings | POTF solved by numerical integration | | | | | | Analytical POTF |
|---|---|---|---|---|---|---|---|
| | $N = 10^1$ | $N = 10^2$ | $N = 10^3$ | $N = 10^4$ | $N = 10^5$ | $N = 10^6$ | 8th-order fitting |
| NMSE | $2.20\times10^{-3}$ | $1.67\times10^{-5}$ | $1.61\times10^{-7}$ | $1.57\times10^{-9}$ | $1.30\times10^{-11}$ | | $1.05\times10^{-10}$ |
| Time (s) | 0.019 | 0.128 | 1.07 | 10.5 | 104 | 1027 | 0.016 |

From these comparison data, it is seen that the numerical integration method produces an accurate POTF only when $N$ is sufficiently large. Concurrently the computational time increases rapidly. By contrast, the polynomial-fitting based analytical calculation method achieves a highly accurate Gaussian POTF in a very short time. For the same accuracy, the numerical integration method requires $N > 10^4$ and a thousand times longer than the analytical method. Further, the analytical calculation method can be implemented by fitting the Gaussian profile using several different polynomials rather than just one. This will improve the POTF's accuracy further since the polynomial fitting performs better in a smaller area. For example, if two 8th-order polynomials are used, the NMSE will drop to $1.88\times10^{-12}$ but the time increases to 0.067s. Even further, if four 8th-order polynomials are used, the NMSE will drop to $1.64\times10^{-13}$ and the time required is 0.13s. In summary, a high-speed, high-accuracy method based on polynomial-fitting for calculating the Gaussian POTF has been established. This tool is now shown to be instrumental in evaluating and optimizing the POTF. The polynomial fitting is not restricted to Gaussian profile, so this method can be generalized to any axially symmetric profile, when appropriate order of polynomial is used.

## 3. POTF Evaluation

### A. Balanced Distribution Criterion

The imaging performance of a QPI system can be predicted by its POTF which specifies the diffracted intensity response distribution across the spatial frequency coverage (SFC). The intensity response, in turn, depends on the profile of the incident illumination. Thus, optimizing this illumination profile is of fundamental importance. A reasonable criterion to achieve overall high-quality imaging is to seek an illumination profile that provides a balanced distribution of the POTF across the entire SFC. With this criterion, all spatial frequencies in the object would be similarly weighted. The criterion for an optimally balanced POTF can be formulated as a minimization. It can be stated as

$$\text{find } S, \text{ min: } \varepsilon = \iint \left[ |T(\eta,\mu)| - |T_0(\eta,\mu)| \right]^2 d\eta d\mu. \tag{10}$$

In this model, $S$ indicates the illumination profile, $T$ is the POTF due to $S$, and $T_0$ is the desired target POTF. Since positive and negative values of POTF produce the same imaging response, absolute values of $T$ and $T_0$ are used in this criterion. For ideal balancing, $T_0(\eta, \mu) = 1$. Due to the complex functional form of $T$, an analytical solution for $S$ is not feasible. Under these circumstances, a discrete optimization can be used to determine a near-optimal $S$. By applying Eq. (4), the optimal illumination $S(k)$ can be divided into a series of uniform disk illuminations with various intensities. In this way, the illumination profile $S(k)$ is expressed as an array $s$ and the corresponding POTF as a matrix. To simplify the representation, the POTF matrices $T$, $T_0$ and $T_U$ are rewritten as column vectors $t$, $t_0$ and $t_U$. The discrete normalized POTF optimization process can then be expressed as

$$\begin{aligned}\text{find } s, \text{ min: } \varepsilon_n &= \sqrt{\frac{\sum(|t|-|t_0|)^2}{M}}, \\ \text{subject to: } t &= \sum_{n=1}^{N} \left( s_n \cdot t_U \left\langle \rho_O, \frac{n}{N} \rho_C \right\rangle \right) = \Omega_{M \times N} s_{N \times 1}, \end{aligned} \tag{11}$$

where $M$ is the number of pixels used in the calculation, and $\Omega$ is the matrix that is spanned by a number of constant POTF vectors $t_U \langle \rho_O, n\rho_C / N \rangle$ of uniform disk illumination. For a given optical system, its normalized numerical aperture $\{\rho_O, \rho_C\}$ determine the SFC. The POTF is non-zero only over this spatial frequency area, and also the calculation of Eq. (11) concerns only this region. Since the POTF is symmetry about the $\eta$-axis and the $\mu$-axis, it is sufficient to consider only the first quadrant in the optimization process. Further, since the POTF varies rapidly near the $\mu$-axis, it is convenient to define a new coordinate system with $\eta' = \eta^2$, $\mu' = \mu$ to enhance the robustness of the POTF evaluation. The coordinate transform is illustrated in Fig. 1(a). In the present optimization, the discrete POTF defined over $(\eta', \mu')$ is used.

## B. Relaxed Optimal Illumination Function

Due to the large number of possible illumination profiles and the nonlinear nature of the criterion, it is not feasible to determine an optimal profile $s$ that makes $\varepsilon_n$ minimal in Eq. (11). However, the criterion can be converted to workable linear optimization problem by removing the absolute value requirement. Thus

$$\text{find } s, \text{ min: } \varepsilon_n \approx \sqrt{\frac{\sum(\Omega s - t_0)^2}{M}} \propto \sum(\Omega s - t_0)^2. \tag{12}$$

The matrix $\Omega$ is known in advance. The optimal solution vector $s$ can be obtained by the method of least square error. Since, in general $M \gg N$, a unique $s$ exists. It is given by

$$s = \left( \Omega^T \Omega \right)^{-1} \Omega^T t_0. \tag{13}$$

Therefore, for any arbitrary target POTF given by $t_0$, a unique illumination profile $s$ can be determined. With the solution vector $s$ thus obtained, the illumination function $S(k)$, in turn, is defined by

$$S(k) = \sum_{n=\lceil Nk/\rho_C \rceil}^{N} s_n, \tag{14}$$

where the symbol $\lceil \; \rceil$ is the round up operation. The accuracy of the solution depends on the discretization precision. A larger $M$ and $N$ contribute to a more accurate solution of the relaxed optimal illumination profile. According to the criterion, a uniform (all-one) POTF vector, $t_0 \equiv 1$, is taken as the target. As shown in Fig. 1(b, c), the relaxed optimal illumination profiles corresponding to the optical configurations $\{\rho_O, \rho_C\}=\{1, 1\}$ and $\{\rho_O, \rho_C\}=\{0.7, 0.5\}$ are solved and drawn, respectively. The POTF has the number of points as $M = 500 \times 500$, where the spatial frequencies $\eta' \in [0, 1]$ and $\mu' \in [0, 2]$ are adopted. A series of $N$ with various values are chosen to demonstrate the robustness of the method, for example $N=\{10, 20, 50\}$. From the drawing of the relaxed optimal illumination profile solutions, it concludes the solved $S(k)$ is stable with $N$. With the increase of $N$, the solved POTF shape approaches the given target shape gradually. Moreover, based on the $N$-$\varepsilon$ curves shown in Fig. 1(b6, c6), $N = 50$ is sufficiently large to solve the optimal illumination and is applied in the subsequent QPI simulations.

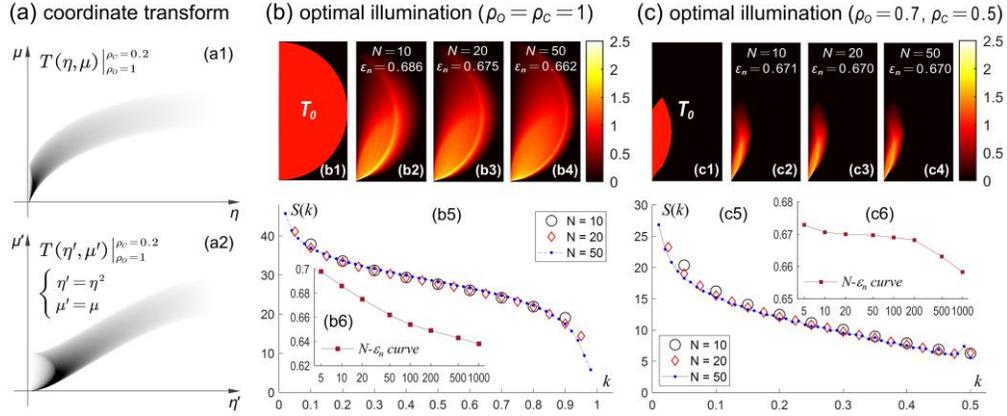

Fig. 1. The relaxed optimal illumination solved by optimizing the POTF under the evaluation rule of Eq. (12). (a) An example to show the coordinate transform used to make robust POTF evaluation. (b) The relaxed optimal illumination profiles and corresponding POTFs solved for a series of $N$ under the optical configuration $\{\rho_O, \rho_C\}=\{1, 1\}$. (c) The relaxed optimal illumination profiles and corresponding POTFs under $\{\rho_O, \rho_C\}=\{0.7, 0.5\}$.

### C. Optimal Gaussian Illumination by Tree Search

The "relaxed optimal illumination" solved by the least square method is not strictly correct because a relaxation of removing absolute symbol was applied. Further, the solved relaxed optimal illumination profile is not constrained in format. As a result, the determined irregular profiles that are drawn in Fig. 1 do not represent readily physically realizable profiles. On the other hand, Gaussian profiles, by comparison, are widely used in practice and have already been shown to be improve imaging accuracy in 3D QPI [9, 30]. In this section, the characteristics of an optimal Gaussian profile are investigated in terms of least square errors. In contrast to the "relaxed optimal illumination" that is solved analytically, the "optimal Gaussian illumination" is determined through tree search in terms of the Gaussian width parameter $\sigma$. All possible values of $\sigma$ are searched to find the optimal Gaussian illumination according to the criterion

$$\text{find } \{\sigma, \xi\}, \text{ min: } \varepsilon = \iint \left[\left(\xi |T_\sigma(\eta', \mu')| - |T_0(\eta', \mu')|\right)\right]^2 d\eta' d\mu', \quad (15)$$

where $\xi$ is a variable that controls the illumination flux. The uniform target POTF is set as unity in this expression, $T_0 = 1$. Compared to Eq. (12), the least square error here is more rigorously defined. Previously, the absolute value representation causing the problem to be nonlinear and thus eliminating the possibility of an analytical solution for $s$ in Eq. (13). However, such a shortcoming does not exist in the case of Eq. (15) since the solution for $\sigma$ is based on a tree search. Simplify the integration into a normalized numerical form, which gives

$$\text{find } \{\sigma, \xi\}, \text{ min: } \varepsilon_n = \sqrt{\frac{\sum(\xi|T_\sigma|-1)^2}{M}}, \quad (16)$$

The summation is over all coordinates ($\eta'$, $\mu'$). Due to the axial symmetry, only the first quadrant ($\eta > 0$, $\mu > 0$) needs to be considered. According to the least square method, at first the solution of $\xi$ is obtained as $\xi = \sum|T_\sigma|/\sum|T_\sigma|^2$. Then a large-scale tree search for the value of $\sigma$ that minimizes $\varepsilon_n$ is performed. The $\sigma$-$\varepsilon$ curves resulting from such searches are shown in Fig. 2(a), where two representative example cases $\{\rho_O, \rho_C\}=\{1, 1\}$ and $\{\rho_O, \rho_C\}=\{0.7, 0.5\}$ are presented. It is seen the optimal Gaussian width $\sigma$ for the two optical configurations are 3.6 and −4.7, respectively. The optimal Gaussian illumination profiles and corresponding POTFs are shown in Fig. 2(b). Note that the scale is not of sense but the trend is critical. The $\varepsilon_n$ values of optimal Gaussian illumination is a little larger than that of the relaxed optimal illumination presented in Fig. 1, which shows the consequence of the constraint on the illumination format. Despite of a larger $\varepsilon_n$, the "optimal Gaussian illumination" guarantees a practical illumination and smooth POTF.

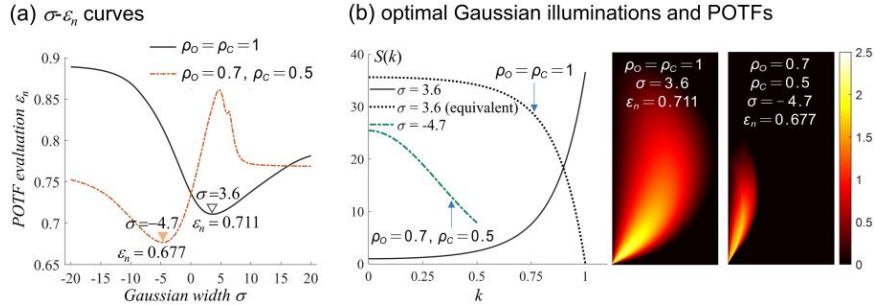

Fig. 2. The optimal Gaussian illuminations obtained by traversal searching under the concept of least square error. (a) $\sigma$-$\varepsilon_n$ curves, from which the optimal $\sigma$ are determined as 3.6 and −4.7 for $\{\rho_O, \rho_C\}=\{1, 1\}$ and $\{\rho_O, \rho_C\}=\{0.7, 0.5\}$, respectively. (b) optimal Gaussian illumination profiles and corresponding POTF distributions.

## 4. Simulation

### A. Simulation Approach

In this section, numerical simulations are presented comparing the phase imaging accuracy of several illumination formats, including uniform illumination, Gaussian illumination, and the "relaxed optimal illumination" proposed in the present work. The simulation is based on the split-step beam propagation method (SSBPM) [9], which provides us with an efficient approach to simulate the 3D diffracted intensity data resulting from illuminating a 3D thick phase object by an incoherent light source. The fundamental principle of SSBPM is that the 3D phase object $n(x, y, z)$ is divided discretely into a series of 2D uniform-RI phase slices $\varphi(x, y)$. Considering only a single point source from the lamp, the emitted lightwave may be regarded as spatially coherent. As this lightwave propagates through each thin slice, its phase is altered and these changes are accumulated to produce the diffracted field $E(x, y, z)$. The diffracted fields due to all of the other lamp point sources are obtained in the say way. Since each point source is incoherent with respect to each other, the total diffracted intensity is the sum of the individual point source diffracted intensities. Thus,

$$I(x,y,z) = \sum_{j=1}^{m} |E_j(x,y,z)|^2, \tag{17}$$

where $m$ is the number of lamp point sources. For a phase-only (non-absorbing) object with an RI distribution close to surrounding medium, $|n - n_{ob}| \ll 1$, the diffraction intensity $I(r)$ and the scattering potential $v(r)$ are linearly related based on the first-order Born approximation. Thus,

$$I(\mathbf{r}) = B + h(\mathbf{r}) * v(\mathbf{r}), \tag{18}$$

where $B$ is the uniform background intensity that is taken to be the average of the measured 3D diffracted intensity $I(r)$, and $h(r)$ is the system phase point-spread function (PSF). The scattering potential $v(r)$ is

$$v = \kappa^2 (n^2 - n_{ob}^2) \approx 2\kappa^2 n_{ob} \Delta n, \tag{19}$$

where $\kappa$ is the wavevector magnitude of the lightwave, $\kappa = 2\pi/\lambda$, $n_{ob}$ is the RI of the surrounding medium and $\Delta n$ is the relative RI of the object, $\Delta n = n - n_{ob}$. It is convenient to recover the object RI in spatial frequency domain. Taking the 3D Fourier transform of both sides of Eq. (18) gives

$$I(\mathbf{f}) = T(\mathbf{f}) \cdot v(\mathbf{f}) \approx 2\kappa^2 n_{ob} T(\mathbf{f}) F[\Delta n(\mathbf{r})], \tag{20}$$

where $\mathbf{f} = (\eta, \mu_x, \mu_z)$, $I(f)$ is the Fourier transform of the diffracted intensity after subtracting the background intensity, $T(f)$ is the system phase optical transfer function (POTF), and the operation $F$ refers to Fourier transform. For a given optical configuration and illumination format, the system POTF can be calculated through the integral method given in [33]. Here, the system POTF is obtained by simulation based on SSBPM. To calibrate the system POTF, a "delta object" is used.

$$\Delta n_\delta(x,y,z) = \begin{cases} t, & x = y = z = 0 \\ 0, & \text{otherwise} \end{cases} \tag{21}$$

where $t$ is a suitably small value (for example, $t = 10^{-6}$) to satisfy the first-order Born approximation. After obtaining the 3D diffraction intensity $I_\delta(f)$ by SSBPM, the system POTF, $T(f)$, can be obtained using $F(\Delta n_\delta) = t$ giving

$$T(\mathbf{f}) = \frac{I_\delta(\mathbf{f})}{v(\mathbf{f})} = \frac{I_\delta(\mathbf{f})}{2\kappa^2 n_{ob} t}. \tag{22}$$

Then, SSBPM is used to generate the 3D diffracted intensity $I(r)$ of the test phase object defined by the relative RI distribution $\Delta n(x, y, z)$. With $I(f)$, the Fourier transform of $I(r)$, the object can then be reconstructed as

$$\Delta n(\mathbf{r}) = F^{-1}\left(\frac{I(\mathbf{f})}{2\kappa^2 n_{ob} T(\mathbf{f})}\right) = F^{-1}\left(t \frac{I(\mathbf{f})}{I_\delta(\mathbf{f})}\right). \tag{23}$$

Note that the system POTF is intrinsically a purely imaginary and odd function. And therefore, it can be proven that the recovered $\Delta n$ in Eq. (23) is purely real. For a given optical configuration and illumination format, the system POTF is invariant so it only needs to be calibrated once using Eq. (22). The application of various illumination formats redistributes the system POTF and therefore can improve the accuracy of object recovery. This will be discussed in detail in following sub-sections. To summarize, the simulation procedure for reconstructing a given 3D phase object consists of four steps:

1) Apply SSBPM to generate the diffracted intensity $I_\delta(r)$ of a "delta object" as defined by Eq. (21).
2) Determine the system POTF using Eq. (22).
3) Apply SSBPM to generate the diffracted intensity $I(r)$ of the unknown 3D phase object.
4) Recover the object RI $\Delta n$ using Eq. (23).

In the calculation of Eq. (23), very small values of the POTF will cause large errors. Therefore, Wiener filtering is applied to make the calculation more robust,

$$\frac{I(f)}{T(f)} \longrightarrow \frac{I(f)T^*(f)}{T(f)T^*(f)+\tau\left(T(f)T^*(f)\right)_{\max}}. \tag{24}$$

where the subscript '$_{\max}$' means the maximum value of the matrix $T(f)T^*(f)$. In addition, Eq. (23) shows that a symmetric relationship exists between the object's RI and the diffracted intensity in 3D QPI based on POTF theory. As shown below, the ratio of the Fourier transforms of the relative RIs of two objects is equivalent to the ratio of the Fourier transforms of the diffracted intensities of these two objects.

$$\frac{F\left[\Delta n_1(r)\right]}{F\left[\Delta n_2(r)\right]} = \frac{F\left[I_1(r)\right]}{F\left[I_2(r)\right]}. \tag{25}$$

In many practical 3D QPI applications, tomographic imaging techniques are adopted to acquire an isotropic spatial resolution. The most significant drawback of non-tomographic QPI is that the "missing cone" problem severely degrades the object reconstruction along the optical axis. This is an inherent shortcoming of non-tomographic phase imaging and cannot be solved by merely increasing the optical $NA_C$ and $NA_O$. In tomographic QPI, either the illumination source or the test object is rotated along an axis that is perpendicular to the optical axis. In this way, the previous missing object spatial frequencies can be obtained. Object rotation provides a simple tomographic imaging technique that uses a stationary optical configuration. In this study, object rotation-based tomographic QPI simulations are performed to investigate the RI recovery accuracy for different illumination formats. According to previous research [9], a rotation angle range of 180° is adequate to collect all possible object spatial frequencies if the RI object is non-absorbing. The number of rotations $R$ is an important parameter that controls the distribution of the synthesized SFC. A suitable choice of $R$ depends on the optical configuration, i.e., $NA_C$ and $NA_O$. For tomographic phase imaging, there are a series of pairs of system POTFs and diffracted intensities measured. The corresponding system POTF and diffracted intensity of the object with a rotation angle $j\alpha$ are pre-processed. The step angle is $\alpha$ and the number of object rotations is $j$. Starting with $j = 1$, the system POTF for $j^{th}$-orientation can be obtained by rotating the system POTF by $-(j-1)\alpha$, meanwhile the diffracted intensity $I(r, j\alpha)$ needs to be rotated back by the same angle to ensure data alignment. The aligned system POTF and the measured intensity corresponding to the $j^{th}$-object are $T_j(f)$ and $I_j(f)$, respectively. Then using the least-squares-method given by Eq. (20a) and Eq. (20b) in [9], a good recovery of the object's RI can be obtained using Eq. (23) after updating Eq. (24) using the synthesized POTF shown below. Equal-angle rotations are treated here, where the angle step is 180°/$R$.

$$\frac{I(f)}{T(f)} \longrightarrow \frac{\sum_{j=0}^{R}\left[I_j(f)T_j^*(f)\right]}{\sum_{j=0}^{R}\left[T_j(f)T_j^*(f)\right]+\tau\left(\sum_{j=0}^{R}\left[T_j(f)T_j^*(f)\right]\right)_{\max}}. \tag{26}$$

Normalized mean square error (NMSE) is used here to evaluate the accuracy of phase imaging. Given an input image $X$ and a recovered image $Y$, the calculation formula of NMSE is given by the following formula, where $X$ with a bar means the average of $X$. Note that it is not possible to obtain the absolute value of object RI $\Delta n$ by POTF-based 3D QPI, since the POTF value at zero spatial frequency is zero and thus the average value of $\Delta n$ is undetermined. However, this does not affect the relative RI distribution of the object, which is generally of more interest than the absolute value of $\Delta n$. In consideration of this point, an offset calculation is used to provide a meaningful accuracy evaluation.

$$NMSE = \frac{\sum(Y - X - (\bar{Y} - \bar{X}))^2}{\sum(Y - \bar{Y})^2}. \tag{27}$$

B. Non-tomographic 3D QPI Test

Considering that the POTF discussed in this study is axially-symmetric and thus can be well expressed by its 2D section profile, the original 3D POTF can be obtained by rotating the 2D POTF around $\eta$-axis. Based on this symmetry, the 3D RI recovery simulation is simplified to a 2D recovery simulation involving a 2D POTF $T(\eta, \mu)$ and a 2D RI object $\Delta n(y, z)$. The object is defined by a grey-scale image that located in the $y$-$x(z)$ plane. Without loss of generality, four images are chosen as the 2D RI object example. As listed in Fig. 3, these RI objects vary in structural complexity but are similar in RI values. The RI range is set to 0.002, $\Delta n \in [0, 0.002]$, which is a suitable value for cell imaging and optical fiber characterization. The first object 'dog' has a very complex structure and the second object 'blocks' exhibits a very simple geometric structure. The third object 'USAF' is used to investigate qualitatively the phase imaging accuracy. For orientational comparison, the fourth object is 'USAF' rotated by 90° to be vertical, named as 'USAF-V', is used to investigate the phase imaging performance of placing the same object in different direction.

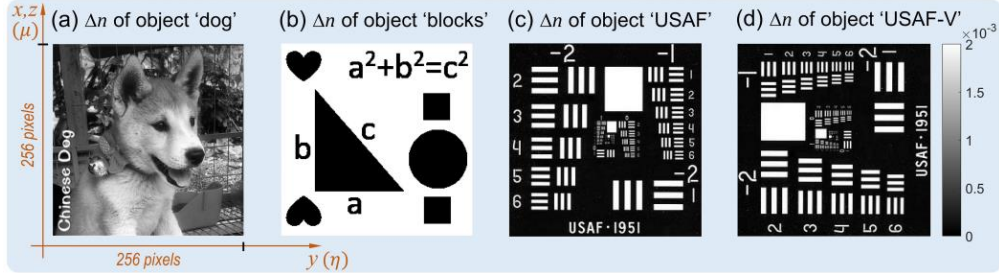

Fig. 3. Four 2D RI objects taken as examples in the POTF-based object reconstruction simulation. The objects $\Delta n(y, z)$ are specified in the $y$-$z$ plane where $y$ is the optical axis. The size of the objects is 256 × 256 pixels. Each pixel has a physical width $\lambda/4$ and therefore the whole object takes $64\lambda \times 64\lambda$ in physical space.

In this section, non-tomographic object reconstruction is simulated under various illumination formats. Since no tomographic rotation used, the SFC is limited and the severe "missing cone" problem occurs along the optical axis. As a result, the structural details in the axial direction are not recovered. In order to mitigate this degradation, the SFC should cover as many spatial frequencies as possible. Toward this goal, the optical configuration is chosen to have $NA_O = NA_C = 1$, $n_{ob} = 1$, $\lambda = 550$ nm. The illumination formats applied in this simulation consist of the "relaxed optimal illumination" and numerous representative Gaussian illuminations, i.e., $\sigma = \{-5, -4, -3, -2, -1, -0.5, -0.1, -0.01, 0.1, 0.5, 1, 2, 3, 3.6, 4, 5\}$. Among them, the "optimal Gaussian illumination" presented in Sec. 3.2 is $\sigma = 3.6$. It is worth mentioning that the system POTF of uniform illumination is all-zero under the current optical configuration $NA_O = NA_C = 1$. A near-uniform Gaussian illumination ($\sigma = -0.01$) is a good approximation to uniform illumination, which helps predict the phase imaging accuracy for uniform illumination. In addition, illumination defined by a positive value of $\sigma$, $\sigma = 1$ for example, is also regarded as Gaussian illumination though it exhibits an inverse-Gaussian profile. All the illumination formats are drawn in Fig. 4(a). Given the illumination and the object's RI, the diffracted intensity $I(r)$ is simulated by SSBPM, where the step size of object slices is set as $\Delta y = \lambda/4 = 134$ nm (a value equivalent to the pixel width). The illumination source is discretized by an interval of $\kappa/64$ in spatial frequency domain. A delta object described by Eq. (21) with $t = 10^{-6}$ is used as an input and the resulting system POTF $T(f)$ is calculated using Eq. (22). The system POTFs for various illumination formats are given in Fig. 4(b) with the same normalization method. It is seen the system POTF for "relaxed optimal illumination" has a large SFC and shows a comparatively balanced distribution between low and high spatial frequencies. This is consistent with the optimization criterion in Sec. 3.1. On the contrary, the system POTF of Gaussian illumination is very smooth but less balanced. A larger $|\sigma|$ represents a rapidly changing illumination that enhances the phase imaging efficiency and produces a more concentrated POTF with a larger range of values. A smaller $|\sigma|$ represents a more uniform illumination. Specifically, the Gaussian illumination with $\sigma = -0.01$ is close to uniform illumination, of which the POTF has a small range of values. The SSBPM is then applied to generate the diffracted intensity $I(r)$ of the four objects. Subsequently their RI maps are reconstructed using Eq. (23), where a Weiner filter with $\tau = 0.02$ is applied in Eq. (24).

The non-tomographic RI recoveries of different objects under various illumination formats are listed in Fig. 4(c). By inspection, the "relaxed optimal illumination" produces top-quality reconstructions for all four objects for the specified illumination formats. For example, the 'dog' object recovered with "relaxed optimal illumination" exhibits more structure details and less stripe artifacts than the Gaussian illuminations. Similar phenomena are seen in the RI recoveries of other test objects. The reason is the POTF for the "relaxed optimal illumination" is less affected by the "missing cone" problem. This explanation is corroborated by the POTF distributions given in Fig. 4(b), where the POTF for the "relaxed optimal illumination" has non-negligible values in the region near the optical axis ($\eta = 0$) while the Gaussian illumination does not. The influence of the "missing cone" problem on the RI recovery can be seen more clearly in the recovered 'USAF' and 'USAF-V' images. For 'USAF', vertical bars are not reconstructed for Gaussian illumination with negative $\sigma$. By comparison, these vertical bars are reconstructed better when "relaxed optimal illumination" is used. Similar results are found for the horizontal bars in the RI recovery of 'USAF-V', where all results are rotated 90° back into horizontal to provide a convenient comparison to 'USAF'. The object reconstructions of 'USAF' and 'USAF-V' verify the advantage of the "relaxed optimal illumination" over typical Gaussian illumination. In addition, it is interesting to see from Fig. 4(b) that the RI recoveries for the near-uniform Gaussian illumination with $\sigma = -0.01$ are obviously worse than for other illuminations. A predictable conclusion is therefore drawn that uniform illumination is not a good choice for non-tomographic high-resolution 3D POTF-based QPI that applies to large $NA_O$ and $NA_C$. Another interesting finding is that the RI recoveries of 'USAF-V' appear better than that for 'USAF', which illustrates the importance object orientation in phase imaging accuracy. The NMSE of the RI recoveries for different illumination formats is presented in Fig. 4(d). The "relaxed optimal illumination" has the smallest NMSE in reconstructing all four RI objects. This demonstrates the generality of the proposed "relaxed optimal illumination" in high-accuracy RI recovery compared to conventional widely-used uniform illumination or Gaussian illumination. The Gaussian illumination with $\sigma = 3.6$ (an optimized value solved by the least square method) achieving good RI recoveries proves the significance of a balanced POTF distribution. In summary, of the illuminations considered, the "relaxed optimal illumination" produces the highest-resolution non-tomographic 3D QPI reconstruction.

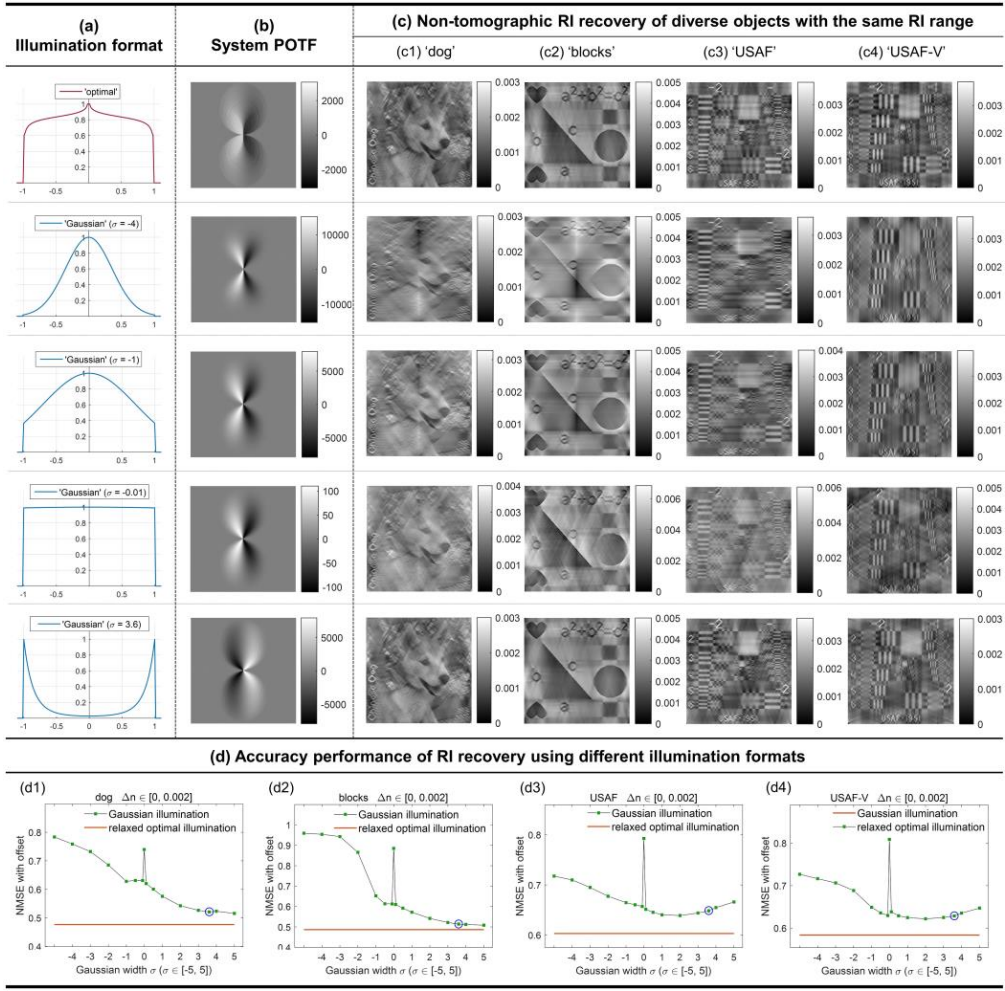

Fig. 4. Non-tomographic object reconstructions under various illumination formats. The optical configuration is $NA_O = NA_C = 1$, $n_{ob} = 1$, $\lambda = 550$ nm. The inputted 2D RI objects are given by Fig. 3. (a) Illumination format, including the "relaxed optimal illumination" and several representative Gaussian illuminations. (b) POTFs for each given illumination, whose scale is normalized by $\kappa^2$. (c) RI recoveries for diverse phase objects under various illuminations. $\tau = 0.02$ is used to supress numerical calculation error. (d) NMSE curves of the RI recoveries. The horizon line on the bottom means the NMSE of the "relaxed optimal illumination", while the circle marks the "optimal Gaussian illumination". The sharp bulges that occur at $\sigma = -0.01$ is a result of severe numerical calculation error since the POTF is very weak.

Then we discuss the phase imaging performance of different illumination formats in recovering a phase object with various RI range. In this simulation, the 'USAF-V' image shown in Fig. 3(d) is chosen as the sample, which has four new RI ranges as {0.001, 0.004, 0.01, 0.02}. The optical configuration and illumination formats are set the same as that in Fig. 4. The illumination formats and corresponding POTFs are given by Fig. 4(a, b) too. SSBPM is applied to simulate the diffracted intensities of 'USAF-V' with various RI ranges under a series of given illumination formats. Then Eq. (23) and Eq. (24) with $\tau = 0.02$ are chosen as the Wiener filter to solve the RI recoveries. The results are listed in Fig. 5(b) and the corresponding NMSE curves are plotted in Fig. 5(c). Still, the "relaxed optimal illumination" provides us with the most accurate object reconstruction among all given illumination formats, and the "optimal Gaussian illumination" with $\sigma = 3.6$ obtains also good reconstructions among all Gaussian illuminations. Therefore, the validity of the proposed "balanced distribution" criterion for POTF evaluation, the correspondingly solved "relaxed optimal illumination" and "optimal Gaussian illumination" in improving phase imaging accuracy are verified once again by the non-tomographic RI recovery test on the USAF-V object with various RI ranges. In addition, as presented in Fig. 5, the phase imaging accuracy drops gradually with the increase of the RI range. The reason lies in the first-order Born approximation. To ensure a good linearization, the test sample must have a small RI range and must be immersed in a suitable RI-matching oil.

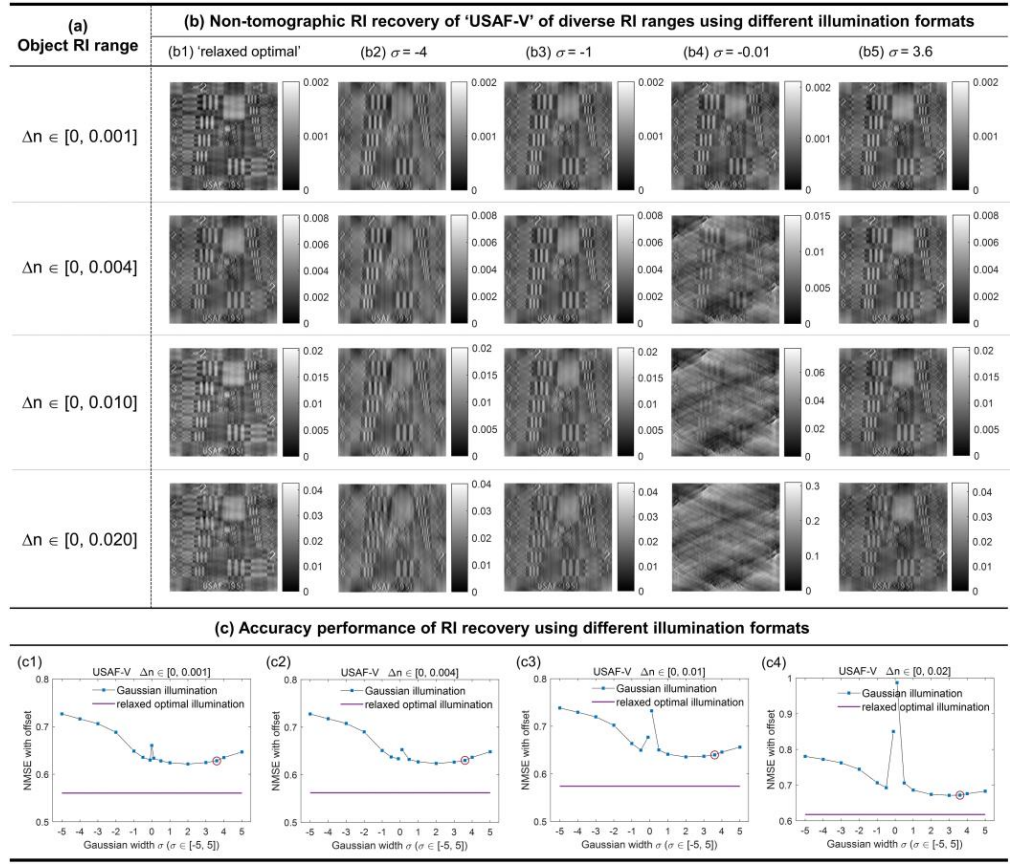

Fig. 5. Non-tomographic object reconstructions of an object with different RI ranges under various illumination formats. The optical configuration is $NA_O = NA_C = 1$, $n_{ob} = 1$, $\lambda = 550$ nm. (a) the list of RI ranges. (b) RI recoveries for the "USAF-V" object with different RI range using various illumination formats. $\tau = 0.02$ is used to supress numerical calculation error. (c) NMSE curves of the RI recoveries. The horizon line on the bottom means the NMSE of the "relaxed optimal illumination", while the circle marks the "optimal Gaussian illumination". Due to a too-large NMSE, NMSEs of $\sigma = -0.01$ are not given in Fig. 5(c3, c4).

## C. Tomographic 3D QPI Test

The tomographic phase imaging simulation takes a more practical optical configuration into account, i.e., $NA_O = 0.7$ and $NA_C = 0.5$. Other simulation settings stay the same, including $n_{ob} = 1$, $\lambda = 550$ nm, $\Delta y = \lambda/4 = 134$ nm, $t = 10^{-6}$ in Eq. (21) and the source discretization step as $\kappa/64$. The test phase object is chosen as 'USAF-V' with a RI range of $\Delta n \in [0, 0.002]$. Since image rotation is needed in tomographic imaging, the 'USAF-V' object needs to shrink until it locates within the maximum circle of the square boundary so that no object structure loses during rotation. Besides, the object size is set as 257×257 pixels by zero-padding to make a better data alignment as odd-size image has a strict center point. The final inputted object, named by 'USAF-S', is shown in Fig. 7(c4). The inner 'USAF' structure takes only 181×181 pixels and the RI on the outer area is set to $\Delta n = 0$. This simulation investigates the tomographic phase imaging accuracy of various illumination formats, including the uniform illumination, the Gaussian illumination and the proposed "relaxed optimal illumination". A series of Gaussian illuminations with diverse $\sigma$ are considered, i.e., $\sigma \in \{-10, -8, -6, -4, -2, -1, -0.5, 0, 0.5, 1, 2, 4\}$. Now the uniform illumination is investigated, and among them it should be mentioned that the "optimal Gaussian illumination" presented in Sec. 3.2 is $\sigma = -4.7$. The tomographic imaging involves 18-angle object rotations ($R = 18$) to generate a synthetized POTF with broad SFC. The choice of $R = 18$ contains other tomography cases such as $R = 9$, $R = 6$, $R = 3$, $R = 2$ and even $R = 1$. The synthetized POTFs of them are presented in Fig. 6, providing us an intuitive impression of the "missing cone" problem in tomographic QPI with different rotations. It is found the severe "missing cone" problem that happens in non-tomographic QPI is improved effectively by applying object rotation. With the increase of $R$, the missing cones of object's spatial frequency are filled well from low-frequency region to high-frequency region gradually. $R = 18$ gives a well-filled SFC so that an isotropic-resolution QPI is achieved probably.

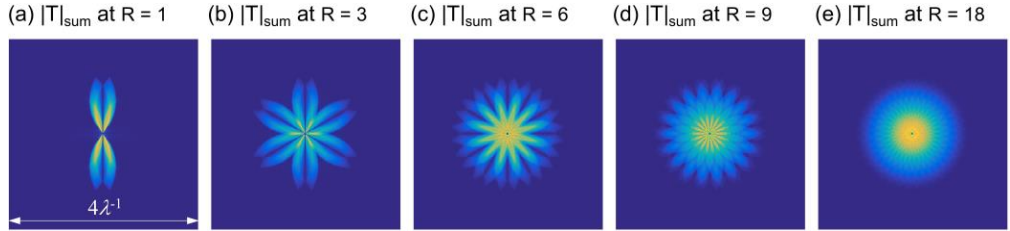

Fig. 6. Synthetized POTFs in tomographic phase imaging with rotation times. Figures shown are the sum of $|T_j(f)|$ where $j$ is integer from 1 to $R$. Optical configuration is set as $NA_O = 0.7$, $NA_C = 0.5$, $n_{ob} = 1$, $\lambda = 550$ nm, and the uniform illumination is taken as the example.

The distribution of the synthetized POTFs as well as the tomographic phase imaging accuracy depends on the illumination format. The illumination formats are listed in Fig. 7(a), and the corresponding basic system POTFs are shown in Fig. 7(b). The tomographic phase imaging results using different number of object rotation in the case of various illumination formats are summarized in Fig. 7(c). The object recoveries get better with the increment of $R$. This is a universal result for all illumination formats. When $R$ is small, the advantage of the "relaxed optimal illumination" over other illuminations is obvious, while this advantage dwindles when $R$ increases. This phenomenon can be seen more clearly from the NMSE curves plotted in Fig. 7(d). Apparently, all NMSE curves appears similar global tendencies and it is found the best choice is always the "relaxed optimal illumination" for all cases of $R$, $R = \{3, 6, 9, 18\}$. Among the Gaussian illuminations investigated, the best one is $\sigma = -4$, which is accordance with the previously presented "optimal Gaussian illumination" ($\sigma = -4.7$) and verifies the significance of the proposed "balanced distribution" concept. However, it is seen clearly from the scale that the improvement on phase imaging accuracy brought by illumination optimization is very limited when a dense-angle tomography is applied. The reason is that the obvious difference on the POTF distribution caused by different illumination profiles is significantly weakened in synthetized POTFs with many rotation angles.

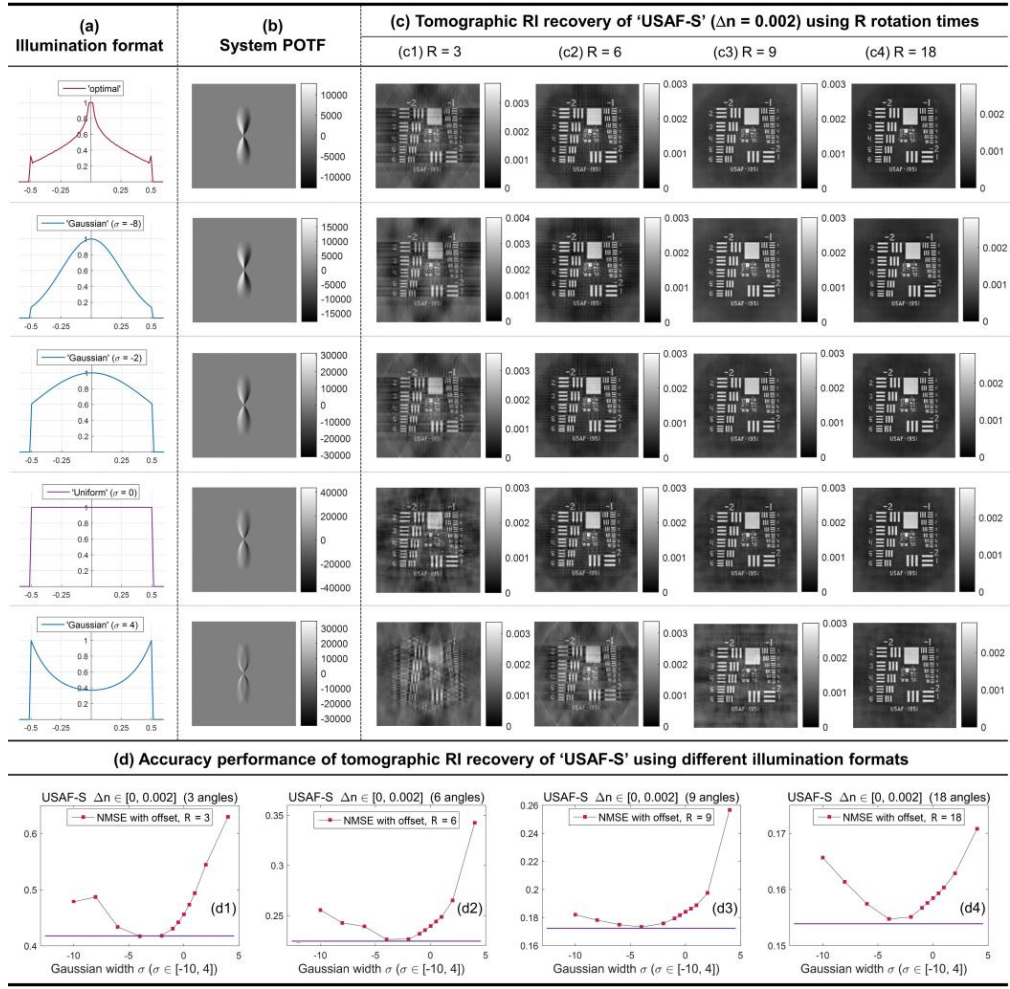

Fig. 7. Tomographic object reconstructions under various illumination formats. The optical configuration is $NA_O = 0.7$, $NA_C = 0.5$, $n_{ob} = 1$, $\lambda = 550$ nm, $\Delta n \in [0, 0.002]$. (a) Illumination formats, including the "relaxed optimal illumination", uniform illumination and several representative Gaussian illuminations. (b) System POTFs for each given illumination formats. (c) RI recoveries for different illumination formats. $\tau = 0.005$ is used to supress numerical calculation error. (d) NMSE curves of the RI recoveries. The horizon line on the bottom is the NMSE of the relaxed optimal illumination.

In summary to the above non-tomographic and tomographic phase imaging simulations, several fundamental results are obtained, including

(1) It is valuable and workable to improve the accuracy of POTF-based 3D QPI by optimizing the illumination format, especially for non-tomographic phase imaging and tomographic phase imaging using few object rotations. And it is insignificant to optimize the illumination format in tomographic phase imaging using many object rotations.

(2) Uniform illumination is not suitable for high-resolution non-tomographic QPI that applies large $NA_O$ and $NA_C$. And generally, the "optimal Gaussian illumination" provides a better RI recovery than uniform illumination, and the "relaxed optimal illumination" provides a best one. These results prove the feasibility of the "balanced distribution" criterion.

## 5. CONCLUSIONS

Using the analytical calculation method based on polynomial fitting developed here, together with the "balanced distribution" criterion for the POTF, a straightforward profile optimization approach has been developed in the present work. This approach has been tested and its effectiveness has been established. Contributions of this work include:

1) An analytical calculation method for the POTF for non-uniform axially-symmetric illumination based on $8^{th}$-order polynomial fitting. This enables very fast and highly accurate calculation of the POTF for Gaussian profiles.

2) The "balanced distribution" criterion for implementing a desired POTF. The resulting POTF is evaluated by square error relative to the desired POTF. Two types of illumination profile solutions are obtained: a "relaxed optimal illumination" profile and an "optimal Gaussian illumination" profile. The "relaxed optimal illumination" profile is obtained analytically from the minimization approach. The "optimal Gaussian illumination" profile is determined by a tree search over all possibilities.

3) Numerical simulations including non-tomographic and tomographic QPI are performed on diverse phase objects with various ranges of RI. These simulation results demonstrate the utility of the "balanced distribution" criterion and the validity of the "relaxed optimal illumination" and the "optimal Gaussian illumination" approaches. These two types of optimized illuminations reveal more high-resolution details in the resulting images than is possible with uniform illumination.

The optimization approach presented in the present work can serve as a basis for further 3D QPI optimization investigations tailored to specific application requirements.

## 6. Funding, acknowledgments, and disclosures

*Funding*

National Science Foundation (NSF) (DGE-1148903 and ECCS-1915971) and Chinese Scholarship Council (CSC).

*Disclosures*

The authors declare no conflicts of interest.

**References**

1. Y. Bao and T. K. Gaylord, "Iterative optimization in tomographic deconvolution phase microscopy," Journal of the Optical Society of America A **35**, 652-660 (2018).
2. D. Jin, R. Zhou, Z. Yaqoob, and P. T. C. So, "Tomographic phase microscopy: principles and applications in bioimaging [Invited]," Journal of the Optical Society of America B **34**, B64-B77 (2017).
3. T. H. Nguyen, C. Edwards, L. L. Goddard, and G. Popescu, "Quantitative phase imaging of weakly scattering objects using partially coherent illumination," Optics Express **24**, 11683-11693 (2016).
4. J. A. Rodrigo, J. M. Soto, and T. Alieva, "Fast label-free microscopy technique for 3D dynamic quantitative imaging of living cells," Biomedical Optics Express **8**, 5507-5517 (2017).
5. J. M. Soto, J. A. Rodrigo, and T. Alieva, "Label-free quantitative 3D tomographic imaging for partially coherent light microscopy," Optics Express **25**, 15699-15712 (2017).
6. C. Zuo, J. Sun, J. Li, A. Asundi, and Q. Chen, "Wide-field high-resolution 3D microscopy with Fourier ptychographic diffraction tomography," Optics and Lasers in Engineering **128**(2020).
7. A. C. Kak and M. Slaney, *Principles of Computerized Tomographic Imaging* (Society for Industrial and Applied Mathematics, Philadelphia, PA. , 2001).
8. Y. Bao and T. K. Gaylord, "Quantitative phase imaging method based on an analytical nonparaxial partially coherent phase optical transfer function," Journal of the Optical Society of America A **33**, 2125-2136 (2016).

9. M. H. Jenkins and T. K. Gaylord, "Three-dimensional quantitative phase imaging via tomographic deconvolution phase microscopy," Applied Optics **54**, 9213-9227 (2015).
10. M. Chen, L. Tian, and L. Waller, "3D differential phase contrast microscopy," Biomedical Optics Express **7**, 3940-3950 (2016).
11. L. Tian and L. Waller, "3D intensity and phase imaging from light field measurements in an LED array microscope," Optica **2**, 104-111 (2015).
12. S. Chowdhury, M. Chen, R. Eckert, D. Ren, F. Wu, N. Repina, and L. Waller, "High-resolution 3D refractive index microscopy of multiple-scattering samples from intensity images," Optica **6**, 1211-1219 (2019).
13. M. Chen, D. Ren, H.-Y. Liu, S. Chowdhury, and L. Waller, "Multi-layer Born multiple-scattering model for 3D phase microscopy," Optica **7**, 394-403 (2020).
14. F. Zernike, "Phase contrast, a new method for the microscopic observation of transparent objects Part I," Physica Scripta **9**, 686-698 (1942).
15. M. R. Teague, "Deterministic phase retrieval: a Green's function solution," J. Opt. Soc. Am. A **73**, 1434-1441 (1983).
16. N. Streibl, "Phase imaging by the transport equation of intensity," Opt. Commun. **49**, 6-10 (1984).
17. C. Zuo, Q. Chen, L. Tian, L. Waller, and A. Asundi, "Transport of intensity phase retrieval and computational imaging for partially coherent fields: The phase space perspective," Opt. Lasers Eng. **71**, 20-32 (2015).
18. C. J. R. Sheppard, "Partially coherent microscope imaging system in phase space: effect of defocus and phase reconstruction," Journal of the Optical Society of America A **35**, 1846-1854 (2018).
19. Y. Bao and T. K. Gaylord, "Multifilter phase imaging with partially coherent light: nonparaxial case," in *Frontiers in Optics*, OSA Technical Digest (online) (Optical Society of America, 2016), JW4A.67.
20. M. H. Jenkins, J. M. Long, and T. K. Gaylord, "Multifilter phase imaging with partially coherent light," Appl. Opt. **53**, D29-D39 (2014).
21. C. Zuo, J. Li, J. Sun, Y. Fan, J. Zhang, L. Lu, R. Zhang, B. Wang, L. Huang, and Q. Chen, "Transport of intensity equation: a tutorial," Optics and Lasers in Engineering **135**, 106187 (2020).
22. C. Zuo, J. Sun, J. Li, J. Zhang, A. Asundi, and Q. Chen, "High-resolution transport-of-intensity quantitative phase microscopy with annular illumination," Scientific Reports **7**, 7654 (2017).
23. X. Ma, Z. Zhang, M. Yao, J. Peng, and J. Zhong, "Spatially-incoherent annular illumination microscopy for bright-field optical sectioning," Ultramicroscopy **195**, 74-84 (2018).
24. X. Ma, B. Zhou, Z. Su, Z. Zhang, J. Peng, and J. Zhong, "Label-free 3D imaging of weakly absorbing samples using spatially-incoherent annular illumination microscopy," Ultramicroscopy **200**, 97-104 (2019).
25. J. Li, A. Matlock, Y. Li, Q. Chen, L. Tian, and C. Zuo, "Resolution-enhanced intensity diffraction tomography in high numerical aperture label-free microscopy," Photonics Research **8**, 1818-1826 (2020).
26. J. Li, Q. Chen, J. Zhang, Y. Zhang, L. Lu, and C. Zuo, "Efficient quantitative phase microscopy using programmable annular LED illumination," Biomedical Optics Express **8**, 4687-4705 (2017).
27. A. Pan, C. Shen, B. Yao, and C. Yang, "Single-shot Fourier ptychographic microscopy via annular monochrome LED array," in *Frontiers in Optics - Proceedings Frontiers in Optics + Laser Science APS/DLS*, 2019),
28. Y. Bao, G. C. Dong, and T. K. Gaylord, "Weighted-least-squares multi-filter phase imaging with partially coherent light: characteristics of annular illumination," Applied Optics **58**, 137-146 (2019).
29. L. Tian and L. Waller, "Quantitative differential phase contrast imaging in an LED array microscope," Opt. Express **23**, 11394-11403 (2015).
30. J. M. Soto, J. A. Rodrigo, and T. Alieva, "Partially coherent illumination engineering for enhanced refractive index tomography," Optics Letters **43**, 4699-4702 (2018).
31. J. Li, Q. Chen, J. Sun, J. Zhang, J. Ding, and C. Zuo, "Three-dimensional tomographic microscopy technique with multi-frequency combination with partially coherent illuminations," Biomedical Optics Express **9**, 2526-2542 (2018).
32. N. Streibl, "Three-dimensional imaging by a microscope," Journal of the Optical Society of America A **2**, 121-127 (1985).
33. J. Huang, Y. Bao, and T. K. Gaylord, "Three-dimensional phase optical transfer function in axially symmetric microscopic quantitative phase imaging," Journal of the Optical Society of America A **37**, 1857-1872 (2020).